\documentclass[10pt]{article}
\usepackage[vmargin=2cm,hmargin=1.5cm,nohead]{geometry}
\usepackage[T1]{fontenc}
\usepackage[francais,english]{babel}
\usepackage{float}
\usepackage[dvips]{graphicx}
\author{\normalsize H. B. RAZAFINDRADINA - P. A. RANDRIAMITANTSOA\\\normalsize Laboratoire LASM\\\normalsize B.P. : 1500, (101) Antananarivo Madagascar\\\normalsize \{hbrazafindradina@gmail.com, rpa@freenet.mg\}}
\title{\normalsize TATOUAGE ROBUSTE ET AVEUGLE DANS LE DOMAINE DES VALEURS SINGULIERES\\ROBUST AND BLIND WATERMARKING IN THE SINGULAR VALUES DOMAIN}
\date{}
\begin{document}
\maketitle
\selectlanguage{francais}
\begin{abstract}
\normalsize Le tatouage d'images consiste a introduire une marque, le nom ou le logo de l'auteur, dans une image dans le but de la prot\'{e}ger contre les copies. La lourdeur de la proc\'{e}dure d'extraction de la marque avec les anciennes m\'{e}thodes de tatouage nous a pouss\'{e} a chercher un nouvel algorithme dans le domaine des valeurs singulieres qui serait aveugle : elle ne n\'{e}cessite pas l'image originale pour extraire la marque. Nous proposons une nouvelle m\'{e}thode robuste qui consiste a ins\'{e}rer les bits de la marque dans la matrice des valeurs singulieres de l'image. Contrairement a la plupart des algorithmes de tatouage, elle est aveugle et les r\'{e}sultats montrent que notre m\'{e}thode est robuste contre la compression JPEG, la r\'{e}duction de couleurs (GIF) ainsi que l'\'{e}talement d'histogramme. Ainsi, nous avons pu obtenir un $PSNR~de~49,63~dB$.\\
Mots-cl\'{e}s : tatouage, images, valeurs singulieres, st\'{e}ganographie, droits d'auteur.
\end{abstract}
\selectlanguage{english}
\begin{abstract}
\normalsize Digital watermarking consists on inserting a mark into an image to protect it against copies. The heaviness of the extraction procedure with the old methods urged us to look for a new algorithm in the singular values domain which would be blind : it does not require the original image to extract the mark. We propose a new robust method which consists on inserting the bits of the mark into the singular values matrix. Contrary to most of the watermarking algorithms, it is blind and the results show that our method is robust against the JPEG compression, the reduction of colors (GIF) and the histogram spreading. So, we were able to obtain $PSNR = 49,63~dB$.\\
Keywords : watermarking, images, singular values, steganography, copyright.
\end{abstract}
\selectlanguage{francais}
\section{Introduction}
Les images constituent la grande partie de l'ensemble des documents num\'{e}riques  manipul\'{e}s et \'{e}chang\'{e}s dans le monde de l'Internet. Il est devenu extrêmement simple de reproduire parfaitement n'importe quel m\'{e}dium. Dans le cas des m\'{e}dia num\'{e}riques (son, image et vid\'{e}o), les recherches s'orientent vers une r\'{e}solution technique en ins\'{e}rant une marque dans le m\'{e}dium afin d'identifier l'ayant droit l\'{e}gitime. Ce m\'{e}canisme d'insertion de marque devrait respecter au moins deux conditions : la marque doit être imperceptible (l'\oe{il} humain ne doit pas pouvoir faire la diff\'{e}rence entre une image marqu\'{e}e et la même non marqu\'{e}e) et robuste (le tatouage doit r\'{e}sister a toutes modifications volontaires ou involontaires). L'extraction devrait être aussi aveugle, c'est-a-dire que pour extraire la marque, on n'a pas besoin de l'image originale.\\
De nombreux algorithmes de tatouage ont \'{e}t\'{e} propos\'{e}s cherchant a optimiser un compromis imperceptibilit\'{e}/robustesse. On peut citer notamment le tatouage dans le domaine DCT, ondelettes et CDMA. Beaucoup de techniques r\'{e}centes de tatouage d'image sont inspir\'{e}es des m\'{e}thodes usuelles de codage et de compression. La d\'{e}composition en valeurs singulieres (SVD) en est un exemple. Cette technique a d\'{e}ja fait ses preuves dans le domaine de la compression \cite{roue} en offrant une qualit\'{e} et un taux de compression proche du JPEG. En ce qui concerne le tatouage SVD, seul un petit nombre de publications existe. Parmi les solutions trouv\'{e}es : la m\'{e}thode de C. Bergman and J. Davidson \cite{bergman} applique la marque dans la matrice $U$, Ruizhen Liu and Tieniu Tan \cite{liu} ont propos\'{e} d'ajouter la marque aux valeurs singulieres $S$ de la matrice en utilisant un poids d'insertion variable, A. Sverdlov, S. Dexter et A. M. Eskiciglu \cite{sverdlov}\cite{ganic} ont propos\'{e} un sch\'{e}ma de tatouage DCT-SVD hybride pour am\'{e}liorer la robustesse. Une approche plus sophistiqu\'{e}e propos\'{e}e par R. Agarwal and M. Santhanam \cite{agarwal} insere la marque dans la matrice $V$ en ajoutant a la matrice $V$ de l'image hôte, la matrice $V$ de la marque. Ces diff\'{e}rents algorithmes sont de type non aveugles.\\
Dans cet article, nous pr\'{e}sentons un nouvel algorithme aveugle qui consiste a ins\'{e}rer la marque dans la matrice $S$ des valeurs singulieres. Les principes fondamentaux de la d\'{e}composition SVD sont tout d'abord rappel\'{e}s, la m\'{e}thode d\'{e}velopp\'{e}e est ensuite d\'{e}taill\'{e}e et les r\'{e}sultats exp\'{e}rimentaux obtenus discut\'{e}s.
\section{M\'{e}thodes utilis\'{e}es}
\subsection{Principe}
Toute matrice $I$ de taille $m\times{n}$ de rang $k$ peut être d\'{e}compos\'{e}e en somme pond\'{e}r\'{e}e de matrices unitaires $m\times{n}$ par SVD.\\
En effet, la d\'{e}composition en valeurs singulieres repose sur le fait qu'il existe \cite{press} une matrice carr\'{e}e $U$ unitaire de taille $m$ et une matrice $V$ unitaire de taille $n$ telles que :
\begin{equation} \label{eq1}
	U^{'}\times{I}\times{V}=S
\end{equation}
Où $S$ est une matrice dont les $r$ premiers termes diagonaux sont positifs, tous les autres \'{e}tant nuls.\\
Les $k$ termes non nuls sont appel\'{e}s valeurs singulieres de $I$. Comme $U$ et $V$ sont unitaires, on a les relations suivantes :
\begin{equation} \label{eq2}
	U\times{U^{'}}=Id(m)
\end{equation}
\begin{equation} \label{eq3}
	V\times{V^{'}}=Id(n)	
\end{equation}
Ainsi :
\begin{equation} \label{eq4}
	I=U\times{S}\times{V^{'}}
\end{equation}
On note $\sigma_{i}$ les valeurs singulieres de $I$ :
\begin{equation} \label{eq5}
	S=
	\left( \begin{array}{ccc}
	\sigma_{1} & \ldots & 0 \\
	\vdots & \ddots & \vdots \\
	0 & \ldots & \sigma_{n}
	\end{array} \right)
\end{equation}
Où $\sigma_{1}\geq{\sigma_{2}}\geq{\dots}\geq{\sigma_{k}}$ et $\sigma_{k+1}=\sigma_{k+2}=\dots=\sigma_{n}=0$\\
Les colonnes de la matrice unitaire $U$ sont en fait les vecteurs propres de la matrice sym\'{e}trique $I\times{I^{'}}$ (les valeurs propres de $I\times{I^{'}}$ \'{e}tant les carr\'{e}s des valeurs singulieres de $I$).
De même, les colonnes de $V$ sont les vecteurs propres de la matrice sym\'{e}trique $I^{'}\times{I}$ (les valeurs propres de $I^{'}\times{I}$ sont les mêmes que celles de $I\times{I^{'}}$).
\subsection{Caract\'{e}ristiques}
Les principales propri\'{e}t\'{e}s de la SVD d'une image sont :
\begin{itemize}
	\item Les valeurs singulieres d'une image ont une tres bonne stabilit\'{e} \cite{liu}, c'est-a-dire, quand une petite perturbation (par exemple une marque) est ajout\'{e}e a une image, les valeurs singulieres ne change pas significativement ;
	\item Les valeurs singulieres repr\'{e}sentent l'\'{e}nergie \cite{roue} de l'image c'est-a-dire que la SVD range le maximum d'\'{e}nergie de l'image dans un minimum de valeurs singulieres.
\end{itemize}
\subsection{Sch\'{e}ma de tatouage SVD}
On utilise la matrice $S$ des valeurs singulieres de l'image hôte pour ins\'{e}rer les bits de la marque.
\begin{itemize}
	\item $I$ repr\'{e}sente l'image originale
	\item $W$ repr\'{e}sente la marque
	\item $I_{w}$ repr\'{e}sente l'image tatou\'{e}e
\end{itemize}
\begin{figure}[h]
	\centering
		\includegraphics[height=3cm,width=7.5cm]{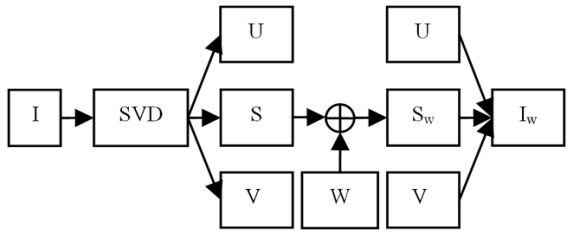}
	\caption{Sch\'{e}ma de tatouage SVD}
	\label{fig1}
\end{figure}
\selectlanguage{francais}
\subsection{Insertion de la marque dans le domaine de la SVD}
Comme dans le cas de la Transform\'{e}e en Cosinus Discrete (DCT), la plus grande partie de l'\'{e}nergie est concentr\'{e}e dans les basses fr\'{e}quences \cite{sverdlov} qui sont exprim\'{e}es par les coefficients les plus proches du coin sup\'{e}rieur gauche de la matrice. C'est cette propri\'{e}t\'{e} qu'on exploite en tatouage : on insere la marque dans les basses ou moyennes fr\'{e}quences selon un compromis robustesse / imperceptibilit\'{e}. C'est pourquoi nous avons choisi le coefficient $\sigma_{3}$ pour ins\'{e}rer la marque.
L'algorithme d'insertion de la marque est d\'{e}crit comme suit :
\begin{enumerate}
	\item Partitionnement de l'image en blocs carr\'{e}s de $8\times{8}$ ;	
	\item On choisit un nombre $b$ de blocs en fonction d'une cl\'{e} qui d\'{e}termine la position de chaque bloc ;
	\item Calcul de la SVD sur chacun des blocs choisis ;
	\item Insertion des bits de la marque dans la matrice $S$ selon les regles suivantes :
\begin{itemize}
	\item On calcule pour chaque bloc :
  \begin{equation} \label{eq6}
    moy=\frac{\sigma_{2}+\sigma_{4}}{2}
  \end{equation}
	\item Pour marquer un \og1\fg, choisir $\sigma_{3}$ telle que : $moy < \sigma_{3} < \sigma_{4}$.
	\item Pour marquer un \og0\fg, choisir $\sigma_{3}$ telle que : $\sigma_{2} < \sigma_{3} < moy$.
	\item On note $S_{w}$ la nouvelle matrice $S$ tatou\'{e}e.
\end{itemize}
	\item Reconstruction de l'image en calculant :
	\begin{equation} \label{eq7}
		I_{w}=U\times{S_{w}}\times{V^{'}}
	\end{equation}
\end{enumerate}
\subsection{Extraction de la marque}
Il suffit de d\'{e}composer chaque bloc en ses valeurs singulieres, puis calculer la moyenne $moy$.\\
Si ($\sigma_{3}>moy$) alors\\
\indent\indent Bit de la marque = \og1\fg\\
Sinon\\
\indent\indent Bit de la marque = \og0\fg\\
Fin si\\
\section{R\'{e}sultats et discussion}
Tous les tests ont \'{e}t\'{e} effectu\'{e}s sur les images \og lena \fg~et \og mandrill \fg~de dimension $512\times{512~bits}$. La marque est une image binare de taille $N = 64~bits$.
\subsection{Imperceptibilit\'{e} : \'{e}valuation de la distorsion}
\begin{figure}[H]
	\centering
		\includegraphics[height=1cm,width=1cm]{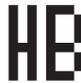}
	\caption{Marque ins\'{e}r\'{e}e 8x8 bits}
	\label{fig2}
\end{figure}

\begin{figure}[!tbp]
	\centering
		\includegraphics[height=5cm,width=5cm]{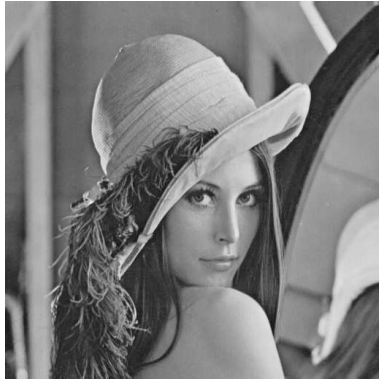}
	\caption{Image lena originale}
	\label{fig3}
\end{figure}

\begin{figure}[!tbp]
	\centering
		\includegraphics[height=5cm,width=5cm]{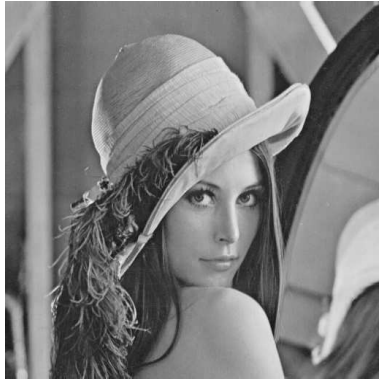}
	\caption{Image lena marqu\'{e}e}
	\label{fig4}
\end{figure}

\begin{figure}[!tbp]
	\centering
		\includegraphics[height=5cm,width=5cm]{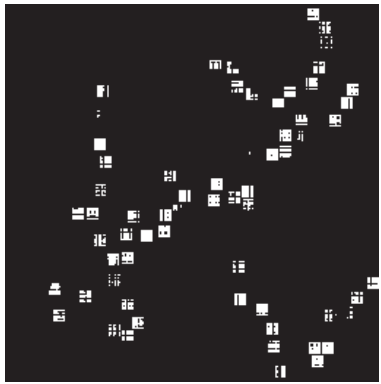}
	\caption{Diff\'{e}rence entre les 2 images lena}
	\label{fig5}
\end{figure}

\begin{figure}[!tbp]
	\centering
		\includegraphics[height=5cm,width=5cm]{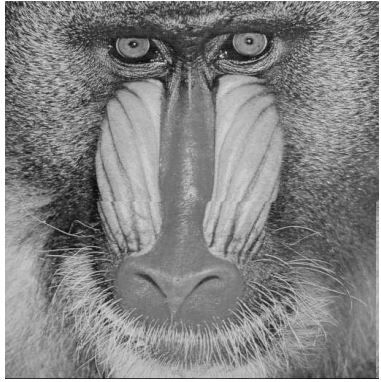}
	\caption{Image mandrill originale}
	\label{fig6}
\end{figure}

\begin{figure}[!tbp]
	\centering
		\includegraphics[height=5cm,width=5cm]{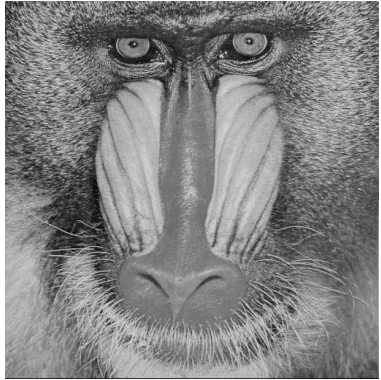}
	\caption{Image mandrill marqu\'{e}e}
	\label{fig7}
\end{figure}

\begin{figure}[!tbp]
	\centering
		\includegraphics[height=5cm,width=5cm]{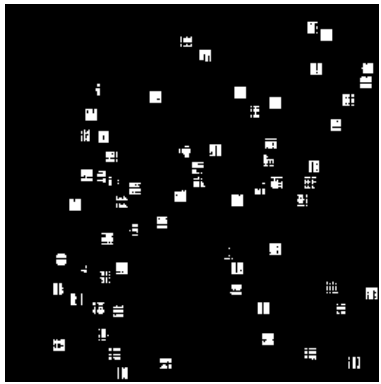}
	\caption{Diff\'{e}rence entre les 2 images mandrill}
	\label{fig8}
\end{figure}

En image, on utilise traditionnellement le PSNR \cite{tamtaoui}\cite{raynal} pour exprimer la distorsion ou l'impact de l'insertion de la marque sur l'image. Soit $I$ l'image originale et $I_{w}$ l'image marqu\'{e}e. Les deux images sont de même taille : $m\times{n}$.
\begin{equation} \label{eq8}
	PSNR=10\times{\log_{10}}\left(\frac{Max\left(I\left(i,j\right)\right)^{2}}{EQM}\right)
\end{equation}
L'EQM est l'erreur quadratique moyenne :
\begin{equation} \label{eq9}
	EQM=\frac{1}{m\times{n}}\sum_{i=1}^m\sum_{j=1}^n\left(I\left(i,j\right)-I_     {w}\left(i,j\right)\right)^{2}
\end{equation}\\
Les figures \ref{fig2}, \ref{fig3}, \ref{fig4}, \ref{fig5}, \ref{fig6}, \ref{fig7}, \ref{fig8} suivantes repr\'{e}sentent la marque ins\'{e}r\'{e}e, les versions originales et tatou\'{e}es des deux images lena.tif et mandrill.tif ainsi que les diff\'{e}rences entre les images non marqu\'{e}es et marqu\'{e}es.\\
La distorsion est \'{e}valu\'{e}e a $PSNR = 49.63~dB$, ce qui correspond a une \og force 10 \fg  \cite{raynal}.\\
\selectlanguage{francais}
\subsection{Robustesse : Evaluation de la r\'{e}sistance aux attaques}
\subsubsection{Objectif}
Id\'{e}alement, \'{e}tant donn\'{e} une image tatou\'{e}e, une entit\'{e} non autoris\'{e}e ne doit pas être capable de d\'{e}truire la marque, c'est-a-dire que la marque doit r\'{e}sister au traitement de signal commun et des attaques intentionnelles.
La robustesse de la m\'{e}thode est \'{e}valu\'{e}e par la similarit\'{e} \cite{tamtaoui} entre la marque originale $W$ et la marque extraite $W'$ en calculant :\\
\begin{equation} \label{eq10}
	 corr(W,W') = \frac{\sum_{i=1}^n(W_{i}-\overline{W})(W'_{i}-\overline{W'})}{\sqrt{\sum_{i=1}^n(W_{i}-\overline{W})^{2}}\sqrt{\sum_{i=1}^n(W'_{i}-\overline{W'})^{2}}}
\end{equation}
où $(W_{i})$ et $(W'_{i})$ sont les composantes respectives de la marque $(W)$ et de la marque extraite $(W')$.
\subsubsection{Condition de robustesse de la m\'{e}thode}
Comme notre m\'{e}thode ins\'{e}re la marque dans les moyennes de coefficients de la SVD, pour assurer sa robustesse, il faut choisir les coefficients $\sigma_{3}$ tels que les \'{e}carts entre les coefficients soient sup\'{e}rieurs a $E$ :
\begin{equation} \label{eq11}
	\sigma_{2} - \sigma_{3} \geq E ~ et ~ \sigma_{3} - \sigma_{4} \geq E
\end{equation}
$E$ : Ecart entre les coefficients, $E_{min} = 64$.\\
Nous avons choisi pour tous les tests $E = 64$. En effet, en choisissant $E < 64$, la m\'{e}thode n'est pas robuste face aux attaques classiques de tatouage (coefficient de corr\'{e}lation tres faible).\\
En augmentant $E$, on peut am\'{e}liorer la robustesse de l'algorithme. \newpage
\subsubsection{Tests}
Nous avons test\'{e} notre technique vis-a-vis des attaques de traitement d'images \cite{petitcolas} :
\begin{itemize}
	\item Compression JPEG~;
	\item Etalement d'histogramme~;
	\item R\'{e}duction de couleurs (GIF)~;
	\item Filtrage m\'{e}dian~;
	\item Ajout de bruits (poivre et sel, gaussien).
\end{itemize}
Le tableau suivant pr\'{e}sente les coefficients de corr\'{e}lation apres d\'{e}tection de la marque pour chaque type d'attaques :\\
\begin{table}[H]
	\centering
	\centering
		\includegraphics[height=1.5cm,width=12cm]{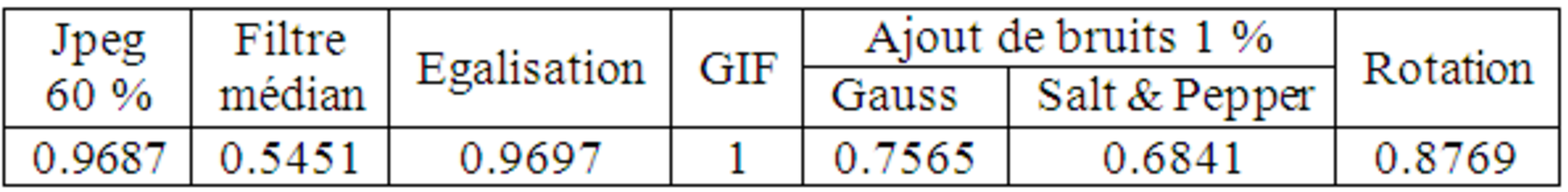}
	\caption{Coefficient de corr\'{e}lation pour chaque type d'attaques}
	\label{tab1}
\end{table}
La figure suivante montre que la m\'{e}thode est robuste jusqu'a une compression correspondant a un facteur de qualit\'{e} de 60 \%. Au-dela de cette valeur, le coefficient de corr\'{e}lation diminue mais ce taux est encore acceptable pour identifier la marque jusqu'a un facteur de qualit\'{e} de 40 \%.\\
\begin{figure}[H]
	\centering
		\includegraphics[height=9cm,width=11.5cm]{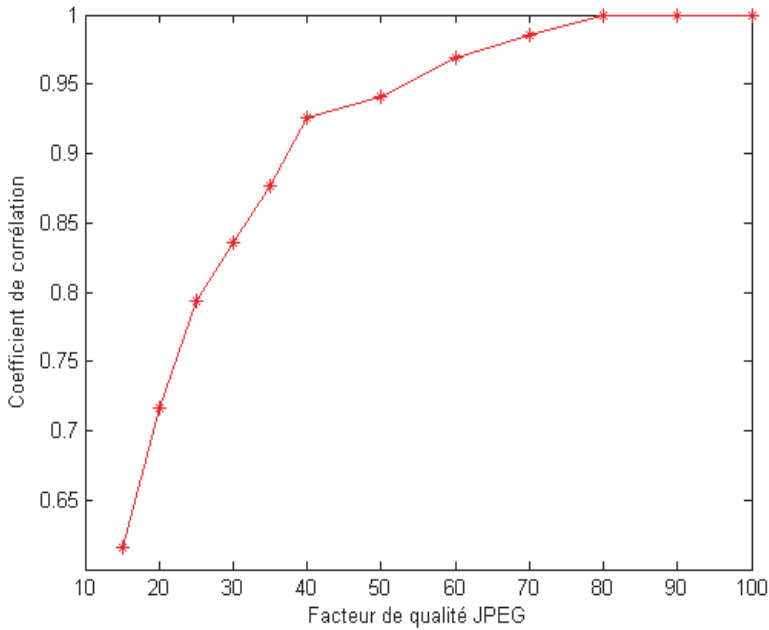}
	\caption{Test de robustesse face a la compression JPEG}
	\label{fig9}
\end{figure}
Ces diff\'{e}rents r\'{e}sultats montrent que la m\'{e}thode est imperceptible et robuste face a 4 types d'attaques (compression, \'{e}talement d'histogramme, r\'{e}duction de couleur, ajout de bruits).\\
N\'{e}anmoins, la m\'{e}thode pr\'{e}sente quelques inconv\'{e}nients tels que :
\begin{itemize}
	\item La m\'{e}thode n'est pas robuste contre le filtrage m\'{e}dian ;
	\item La g\'{e}n\'{e}ration de cl\'{e}s est conditionn\'{e}e par l'ecart $E$, c'est-a-dire on ne prend que les coefficients qui respectent la condition de l'\'{e}quation \ref{eq11}, ce qui entraine la lenteur de l'algorithme de g\'{e}n\'{e}ration de cl\'{e}s ;
	\item L'algorithme de d\'{e}composition en valeurs singulieres n\'{e}cessite un temps de calcul tres important, ce qui p\'{e}nalise la vitesse de la m\'{e}thode.
\end{itemize}
\section{Conclusion et perspectives}
Dans cet article, nous avons propos\'{e} une m\'{e}thode de tatouage \og robuste et aveugle\fg qui ajoute les marques dans les matrices des valeurs singulieres. En ins\'{e}rant les marques dans les moyennes des valeurs singulieres, nous avons pu avoir une technique de tatouage aveugle et robuste contre la compression JPEG, la r\'{e}duction de couleurs, l'ajout de bruits et l'\'{e}talement d'histogramme. Notre m\'{e}thode n'est pas robuste face au filtrage. En variant l'\'{e}cart $E$, on peut contrôler la robustesse de la m\'{e}thode.\\
Nos perspectives se tournent vers la conception d'une m\'{e}thodes SVD qui serait tres robuste contre les attaques g\'{e}ometriques en combinant notre m\'{e}thode avec celle qui utilise un d\'{e}tecteur de points d'int\'{e}rêts \cite{bas}. Des am\'{e}liorations pour augmenter la capacit\'{e} du tatouage sont aussi en cours d'\'{e}tudes.\\
\section{Remerciements}
Nous tenons a remercier le Professeur Driss ABOUTAJDINE de la Facult\'{e} des Sciences de Rabat-Maroc pour ses pr\'{e}cieux conseils et suggestions.\newpage
\selectlanguage{francais}

\end{document}